\documentclass{article}
\usepackage[a4paper, left=2.5cm, right=2.5cm]{geometry}
\usepackage{url}
\usepackage{graphicx}
\usepackage{xcolor}
\usepackage{float}
\usepackage[acronym]{glossaries}
\usepackage{glossaries-extra}
\usepackage{subcaption}
\usepackage{amsmath,amssymb,amsfonts}
\usepackage{rotating}
\usepackage{xspace}
\usepackage{pifont}
\usepackage{booktabs}
\usepackage{multirow}
\usepackage{array}
\usepackage[hidelinks]{hyperref}
\usepackage[%
  bibstyle=ieee,
  citestyle=numeric-comp,
  isbn=true,
  doi=false,
  sorting=none,
  url=true,
  defernumbers=true,
  bibencoding=utf8,
  backend=biber,
  maxnames=2,
  minnames=1
]{biblatex} 

\setabbreviationstyle[acronym]{long-short}

\newcommand{\z}{$Z/\gamma^{*}\rightarrow\tau\tau$\xspace}
\newcommand{\qq}{$Z/\gamma^{*}\rightarrow qq$\xspace}
\newcommand{\tauh}{$\tau_h$\xspace}

\newcommand{\ptvis}{\ensuremath{p_{\mathrm{T}}^{\mathrm{vis}}}\xspace}

\newcommand{\massvis}{\ensuremath{m^{\mathrm{vis}}}\xspace}
\newcommand{\future}{Fu$\tau$ure\xspace}

\newacronym{hps}{HPS}{hadron-plus-strips}
\newacronym{cms}{CMS}{Compact Muon Solenoid}
\newacronym{lhc}{LHC}{large hadron collider}
\newacronym{bdt}{BDT}{boosted decision tree}
\newacronym[longplural={boosted regression trees}, shortplural=BRTs]{brt}{BRT}{boosted regression tree}
\newacronym{dl}{DL}{deep learning}
\newacronym{iqr}{IQR}{interquartile range}
\newacronym{ecal}{ECAL}{electromagnetic calorimeter}
\newacronym[longplural={decay modes}, shortplural=DMs]{dm}{DM}{decay mode}
\newacronym[longplural={recurrent neural networks}, shortplural=RNNs]{rnn}{RNN}{recurrent neural networks}
\newacronym{ild}{ILD}{International Large Detector}
\newacronym{idea}{IDEA}{Innovative Detector for Electron-positron Accelerator}
\newacronym{atlas}{ATLAS}{A Toroidal LHC Apparatus}
\newacronym{tpf}{TPF}{Tau Particle Flow}
\newacronym{clic}{CLIC}{Compact Linear Collider}
\newacronym{fcc}{FCC}{Future Circular Collider}
\newacronym{sota}{SOTA}{state-of-the-art}
\newacronym{cld}{CLD}{CLIC Like Detector}
\newacronym{sm}{SM}{Standard Model of particle physics}
\newacronym{bsm}{BSM}{physics beyond the SM}
\newacronym{nn}{NN}{neural network}
\newacronym[longplural={branching ratios}, shortplural=BRs]{br}{BR}{branching ratio}
\newacronym{pp}{pp}{proton-proton}
\newacronym{ee}{$e^+e^-$}{electron-positron}
\newacronym[longplural={field programmable gate arrays}, shortplural=FPGAs]{fpga}{FPGA}{field programmable gate array}
\newacronym{clicdet}{CLICdet}{CLIC like detector}
\newacronym{bib}{BIB}{beam induced background}
\newacronym{pcgrad}{PCGrad}{Projected Conflicting Gradients}
\newacronym{bce}{BCE}{binary cross entropy}
\newacronym{ce}{CE}{cross entropy}
\newacronym[longplural={long-lived particles}, shortplural=LLPs]{llp}{LLP}{long lived particle}
\newacronym[longplural={heavy neutral leptons}, shortplural=HNLs]{hnl}{HNL}{heavy neutral lepton}
\newacronym[longplural={Large Electron–Positron Collider}, shortplural=LEP]{lep}{LEP}{Large Electron–Positron Collider}

\bibliography{main.bib}

\title{ParticleTransformer is all you need for reconstructing hadronic tau leptons}
\author{
  Seeba, Nalong-Norman$^{1,2}$ \\ \texttt{nalong-norman.seeba@cern.ch} \and 
  Tani, Laurits$^{1,3}$\\ \texttt{laurits.tani@cern.ch} \and
  Lange, Torben$^1$ \\ \texttt{torben.lange@cern.ch} \and
  Pata, Joosep$^1$ \\ \texttt{joosep.pata@cern.ch} \and
}
\date{%
    {\footnotesize $^1$\emph{National Institute Of Chemical Physics And Biophysics (NICPB), Rävala pst. 10, 10143 Tallinn, Estonia}\\
    $^2$\emph{Tallinn University of Technology (TalTech), Ehitajate tee 5, 19086 Tallinn, Estonia}\\
    $^3$\emph{Institute for Experimental Physics, Universität Hamburg, Luruper Chaussee 149, 22761
Hamburg, Germany}}
}

\begin{document}

\maketitle

\begin{abstract}
    The large number of $Z \rightarrow \tau\tau$ events expected during the TeraZ program at FCC-ee will allow for precision measurements and searches for physics beyond the Standard Model, requiring accurate reconstruction of hadronically decaying tau leptons. This reconstruction is particularly challenging due to the presence of undetected neutrinos and the diverse topology of hadronic tau decays, making the design of robust heuristic reconstruction algorithms challenging. In this work, we present the first fully machine learned hadronic tau reconstruction approach tuned for FCC-ee studies. The reconstruction is formulated as a set of complementary tasks, including tau identification, decay mode classification, charge reconstruction, and full four-momentum regression. The algorithms are evaluated on fully simulated electron--positron collision samples with realistic detector effects using the CLD detector setup. We compare dedicated task-specific models with a unified multi-task model and quantify their performance in a granular manner across all reconstruction tasks. Both approaches achieve per-mille-level tau mis-identification rates at high signal efficiency, decay mode classification F1 scores of up to 0.95 for the dominant channels, and sub-per-mille charge mis-identification rates, outperforming a conventional jet-charge estimator by up to two orders of magnitude. For the full kinematic reconstruction, the models achieve per-mille-level angular resolution and percent-level visible transverse momentum resolution, exceeding the performance of reconstruction-level jet observables. While the dedicated training approach achieves the best kinematic reconstruction performance, the unified multi-task approach attains slightly better performance for the classification tasks while using only about one-quarter of the trainable parameters. The resulting models provide a realistic high-performance solution for hadronic tau reconstruction at FCC-ee, offering identification, charge discrimination, decay mode analysis and full kinematic reconstruction.
\end{abstract}

\section{Introduction}

    During TeraZ, the first operational stage of the envisioned \gls{fcc} experiment, approximately $\mathcal{O}(10^{11})$ $\mathrm{Z} \rightarrow \tau\tau$ events are expected to be produced. Making full use of this dataset for precision measurements~\cite{alcaraz_maestre_2025_69a4s-7vw37} and other analyses involving hadronically decaying tau leptons requires accurate \tauh reconstruction.


    However, reconstructing the $\tau$ lepton is inherently challenging. Due to its very short lifetime of only 290 fs, corresponding to an average decay length of $c\tau = 87~\mu\mathrm{m}$, the $\tau$ cannot be observed directly and must be inferred from its decay products. These products always include at least one $\nu_\tau$ that escapes detection, reducing the reconstructable final state. Approximately 1/3 of $\tau$ decays proceed leptonically and 2/3 hadronically. Since leptonic decays are naturally accounted for by electron and muon reconstruction, this work focuses exclusively on hadronically decaying $\tau$ leptons, denoted as \tauh.

    The experimentally accessible decay products define the visible four-momentum ($p^{\mu}$) of the \tauh candidate. Accurately reconstructing the full visible four-momentum of the $\tau_h$ candidate, specifically its transverse momentum, invariant mass, and angular direction, is a fundamental prerequisite for a wide range of precision measurements at future colliders. The reconstruction of the visible transverse momentum ($p_{T}^{vis}$) and mass (\massvis) is critical for determining the true invariant mass of heavy resonances, such as the Higgs and $Z$ bosons, decaying into tau pairs. Because the $\nu_{\tau}$ escapes detection, the visible decay products represent only a fraction of the parent particle's kinematics, resulting in a visible mass distribution that is significantly shifted and broadened~\cite{Elagin:2010aw,Maruyama:2015fis}. Robust visible kinematic regression is therefore the necessary first step for any downstream algorithm aiming to estimate the missing neutrino momenta and recover the true resonance mass~\cite{CMS:2014wdm}. Furthermore, precise angular reconstruction (pseudorapidity $\eta$ and azimuthal angle $\phi$) is essential for both mass approximation algorithms and polarization studies~\cite{alcaraz_maestre_2025_69a4s-7vw37}. Due to the tau lepton's short lifetime, its spin and parity must be inferred from the angular and energy distributions of its visible decay products. In the context of $H \rightarrow \tau\tau$ decays, measuring the angle between the decay planes of the two tau leptons provides sensitivity to the CP structure of the Higgs Yukawa coupling~\cite{Filatov:2023rue}. Finally, many standard mass reconstruction techniques utilize the collinear approximation for highly boosted topologies, which assumes the undetected neutrino is emitted in the same direction as the visible hadronic products; thus, exceptional resolution of the $\tau_h$ direction is necessary~\cite{Filatov:2023rue,Giappichini:2026uly}.

    Beyond kinematic regression, accurate reconstruction of the hadronic tau charge ($q$) is required for both Standard Model measurements and searches for new physics. In di-tau resonance searches (such as $Z \rightarrow \tau\tau$ and $H \rightarrow \tau\tau$), the primary method for reducing combinatorial background is the Opposite-Sign (OS) requirement. Since heavy neutral resonances decay into a $\tau^+$ and a $\tau^-$, selecting candidate pairs with opposite charges suppresses backgrounds from $W$+jets and QCD multijet processes, where misidentified jets have uncorrelated charges~\cite{cms2012performance}. Misidentifying the $\tau_h$ charge causes true signal events to fail this selection, reducing the statistical power of the analysis. Charge reconstruction is also central to the physics program of the \gls{fcc}. At the Tera-$Z$ phase, extracting the electroweak mixing angle ($\sin^2 \theta_{W}^{eff}$) relies on measuring the forward-backward asymmetry ($A_{FB}$) and polarization of tau leptons in $Z$-boson decays~\cite{Dam:2021ibi}. Furthermore, searches for charged Lepton Flavor Violation (cLFV), such as $Z \rightarrow \mu^{\pm}\tau^{\mp}$, depend on charge identification to ensure charge conservation and classify the event topology~\cite{Dam:2018rfz}.

    In general, \tauh reconstruction algorithms approach the problem from one of two directions: directly predicting high-level \tauh quantities (such as Tau ID, decay mode, charge, and four-momentum) from the decay constituents in an end-to-end fashion, or fully reconstructing the individual decay products, leaving the inference of the properties of the \tauh from the decay products to a subsequent model. Classical algorithms, such as the \gls{hps}~\cite{cms2012performance, cms2016reconstruction} at CMS, follow the latter approach, combining charged tracks and calorimeter strips into decay mode specific \tauh candidates and applying isolation and quality cuts for candidate selection. In this paper, we aim to provide the first comprehensive end-to-end model to reconstruct the full \tauh object for future colliders. This approach is intended for analyses that require high-level reconstructed objects for event selection. A full constituent-level reconstruction is required for analyses sensitive to the internal structure of the \tauh decay, such as $\mathrm{H}\rightarrow\tau\tau$~\cite{giappichini_2025_d3y9x-8a819} and $\tau$ polarization studies~\cite{alcaraz_maestre_2025_69a4s-7vw37}, and is not addressed in this paper.

    At present, the state of the art \tauh reconstruction algorithms are based on deep learning and provide important context for this work.
    At CMS, the DeepTau algorithm~\cite{CMS:2022prd} has established a strong benchmark
    for \tauh identification using a deep neural network with convolutional layers, and more
    recently alternative approaches based on graph neural networks and ParticleTransformer architectures have been explored~\cite{Cardini:2025zbp}. A notable development in this direction is the Tau Transformer (TaT), which utilizes an embedding module and self-attention layers to process the multimodality of the input representation, demonstrating significantly improved background rejection over DeepTau~\cite{Filatov:2023rue}. A systematic comparison of DeepTau, ParticleNet~\cite{Qu:2019gqs}, and UParT~\cite{CMS-DP-2024-066} has also been carried out~\cite{CMS-DP-2025-073}. Specialized variants targeting particular physics regimes have furthermore been developed, including DisTau~\cite{CMS-DP-2024-053} for \tauh candidates displaced from the
    primary vertex, Boosted DeepTau~\cite{CMS-DP-2025-047} for highly boosted topologies where the two $\tau$ leptons overlap, and TauNet~\cite{Collaboration:2905110}, a dedicated energy-flow neural network targeting low-$p_T$ \tauh reconstruction in the CMS Run 3 scouting data stream.
    At ATLAS, the \tauh reconstruction and identification employs recurrent neural networks
    (RNNs) for both the identification and electron rejection tasks, as described in Ref.~\cite{ATL-PHYS-PUB-2022-044}. In lepton collider environments, the TauFinder algorithm~\cite{Muennich:1443551,Gallinaro:2026djh}, originally developed for linear colliders~\cite{Muennich:1443551} and more recently applied to muon collider studies, is a notable example of a \tauh reconstruction approach optimized for such conditions. It targets decay modes with one or three charged tracks and applies sequential quality cuts on the reconstructed candidates. While these algorithms mostly address individual reconstruction tasks, the present work aims to provide a unified, transformer-based approach that can be directly deployed for ongoing and upcoming FCC sensitivity analyses.
    
\begin{figure}[H]
            \centering
            \includegraphics[width=\textwidth]{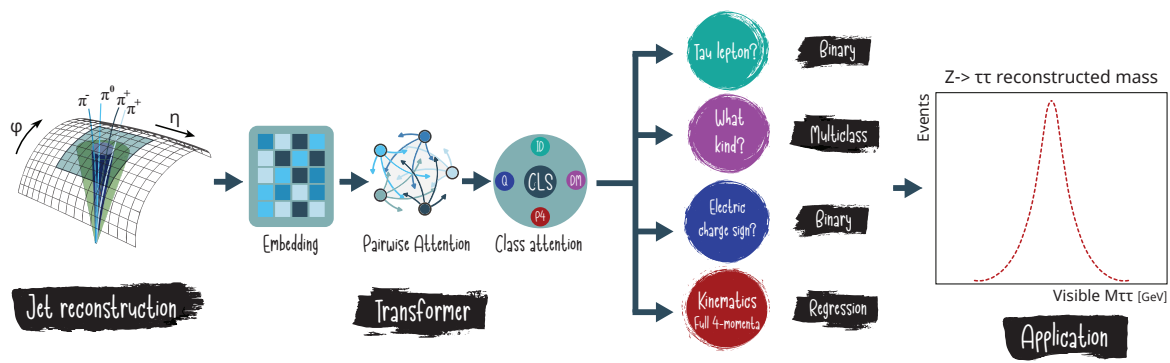}
            \caption{Illustration of \tauh reconstruction. Particle flow candidates forming a reconstructed jet are processed by ParticleTransformer-based models through an embedding network and layers of pairwise particle attention. Task-specific output heads produce four reconstruction targets: \tauh identification, decay mode classification, charge reconstruction, and full four-momentum regression. The resulting \tauh objects can be directly used in downstream physics analyses.}
            \label{fig:abs}
        \end{figure}

    It has been shown previously that the ParticleTransformer~\cite{Qu:2022mxj} -based high-level reconstruction algorithm performs well on the following tasks: \tauh identification~\cite{Lange:2023gbe}, decay mode and $p_T$ regression~\cite{TANI2025109399}. In this work we expand the range of reconstruction tasks to include full 4-momentum regression, tau identification, charge identification and decay mode classification in a single, unified model. In principle, this approach is extensible to additional properties such as the decay vertex~\cite{kuhn1993tau}, needed for studying displaced vertex signatures, for example in analyses featuring \glspl{llp}~\cite{knapen2023guide,lee2019collider,banerjee2018novel} such as \glspl{hnl}~\cite{abdullahi2023present,gronau1984extending}. An illustration of the \tauh reconstruction problem addressed in this paper is shown in \autoref{fig:abs}. Reconstructed jet constituents serve as inputs to ParticleTransformer-based models, producing high-level \tauh objects across four reconstruction tasks suitable for downstream physics analyses at \gls{fcc}-ee .

    We study two approaches, training a dedicated model separately for each of the four tasks, referred to as SingleParTau, and training a unified multi-task model, referred to as MultiParTau, on all four tasks simultaneously. The performance of both approaches is evaluated in a granular way, using task-specific metrics for each of the four reconstruction tasks.

    An overview of the dataset and its features is given in \autoref{sec:dataset}, followed by the introduction to the architecture and training strategies of the two approaches. The results are presented in \autoref{sec:results} and a short summary and thoughts about the next steps are given in \autoref{sec:outlook}.


\section{Dataset}\label{sec:dataset}

    With this paper, we introduce the new version of the jet-based \future dataset for \tauh reconstruction~\cite{dataset}. This dataset aims to align with the current efforts of the envisioned \gls{fcc}~\cite{benedikt2018fcc,fcc2019fcc} experiment. We used the \gls{cld} detector concept~\cite{Bacchetta:2019fmz}, one of the possible detectors at \gls{fcc}, for event generation, simulation and reconstruction.
    
    Furthermore, the dataset is updated to include approximately two million $\mathrm{e}^+\mathrm{e}^-$ collision events with \z (signal) and \qq (background) samples generated at $\sqrt{s} = 91$ GeV, corresponding to the TeraZ run.
    
    The Monte Carlo generation and event reconstruction are described in \autoref{sec:mc-generation} and the input features for the models in \autoref{sec:features}. This public dataset can be found in Ref.~\cite{dataset} and the software used to produce it in Ref.~\cite{data_software}.

    \subsection{Monte Carlo samples and event reconstruction}\label{sec:mc-generation} 

        The $e^{+}e^{-}$ collision events in this dataset are generated using PYTHIA8~\cite{10.21468/SciPostPhysCodeb.8} with default tunes configured via the Key4HEP generation wrapper. The generation assumes a clean environment without beam-induced backgrounds; beamstrahlung and beam energy spread effects are not enabled, since their impact on \tauh reconstruction and identification is expected to be negligible. The decays of the $\tau$ leptons are handled directly by the default PYTHIA8 decay tables, without the use of external polarization packages.

        The subsequent full detector simulation is performed with Geant4~\cite{Agostinelli:2002hh} using the \texttt{FTFP\_BERT} physics list. The detector geometry utilizes the \texttt{CLD\_o2\_v07} setup, an evolution of the CLD detector concept envisaged for the \gls{fcc}-ee~\cite{Bacchetta:2019fmz}. The CLD features an all-silicon vertex and tracking system, highly granular calorimeters (a silicon-tungsten ECAL and a scintillator-steel HCAL), and a superconducting solenoid providing a 2~T magnetic field.

        The simulated events are reconstructed using the Marlin~\cite{Gaede:2006pj} and Key4HEP~\cite{Ganis:2021vgv} software stack (release \texttt{2025-05-29}). Event reconstruction is performed in a clean environment without background overlay. Charged particle tracks are reconstructed using the ConformalTracking algorithm~\cite{Brondolin:2019awm} with optimized settings for the \texttt{CLD\_o2\_v07} geometry, and prepared in the EDM4HEP~\cite{Gaede:2022leb} format. The features used by the \tauh identification and reconstruction model(s) are extracted from the particle flow candidates reconstructed by PandoraPF~\cite{Marshall:2012ry,Marshall:2015rfa}.

        All generator cards, steering configurations, and PandoraPF calibration constants are managed via the dedicated CLD configuration repository~\cite{leonhard_reichenbach_2026_20541598}.


    \subsection{Input features and validation}\label{sec:features}

        The dataset is based on jets clustered from particle flow candidates reconstructed by PandoraPF. Jets are clustered using the \texttt{ee\_genkt}~\cite{Boronat:2016tgd} algorithm with $\mathrm{p}=-1$, $\mathrm{R}=0.4$ and with no minimum $p_T$ cut applied. The cut on $p_T$ is removed in this version in order to retain jets matched to $\tau_h$ candidates with $p_T$ of a few GeV, relevant for the TeraZ physics program.

        For each charged particle flow candidate, the track parameters $\mathrm{d}_{\mathrm{xy}}$, $\mathrm{d}_{\mathrm{z}}$, $\mathrm{S}_{\mathrm{d}_{\mathrm{z}}}$ and $\mathrm{S}_{\mathrm{d}_{\mathrm{xy}}}$ are computed using the track parametrization of Ref.~\cite{Kramer:2006zz}. The corresponding values for neutral particle flow candidates are set to dummy values to maintain a consistent feature set. The complete list of features used in the training is listed in \autoref{tab:partau-features}.

        \begin{table}
        \centering
        \caption{Jet constituent (particle flow candidate)-level input features used in the model.}
        \label{tab:partau-features}
            \begin{tabular}{lll}
            \toprule
            \multicolumn{3}{c}{\textbf{Constituent kinematics}} \\
            \midrule
            \midrule
            \textbf{Category} & \textbf{Feature} & \textbf{Description} \\
            \midrule
            \multirow{4}{*}{Kinematics (absolute)}
            & $p_x$ & Candidate momentum x-component \\
            & $p_y$ & Candidate momentum y-component \\
            & $p_z$ & Candidate momentum z-component \\
            & $E$   & Candidate energy \\
            \midrule
            \multicolumn{3}{c}{\textbf{Constituent features}} \\
            \midrule
            \midrule
            \textbf{Category} & \textbf{Feature} & \textbf{Description} \\
            \midrule
            \multirow{7}{*}{Kinematic} 
            & $\log p_T$ & Log transverse momentum \\
            & $\log E$   & Log energy \\
            & $\log (p_T^{\mathrm{rel}})$ & Log relative $p_T$ w.r.t. jet \\
            & $\log (E^{\mathrm{rel}})$   & Log relative energy \\
            & $\Delta \eta$ & Relative pseudorapidity from the jet axis \\
            & $\Delta \phi$ & Relative azimuthal angle from the jet axis \\
            & $\Delta R$    & Angular distance to jet axis \\            
            \midrule
            
            \multirow{6}{*}{Particle ID} 
            & isElectron & Electron indicator \\
            & isMuon & Muon indicator \\
            & isPhoton & Photon indicator \\
            & isChargedHadron & Charged hadron indicator \\
            & isNeutralHadron & Neutral hadron indicator \\
            & charge & Candidate charge \\
            
            \midrule
            
            \multirow{4}{*}{Impact parameters} 
            & $d_z$ & Longitudinal impact parameter \\
            & $\mathrm{S}_{\mathrm{d}_{\mathrm{z}}} =  \frac{\mathrm{d}_{\mathrm{z}}}{\sigma(\mathrm{d}_{\mathrm{z}})}$ & Significance of the longitudinal impact parameter $d_z$ \\
            & $d_{xy}$ & Transversal impact parameter \\
            & $\mathrm{S}_{\mathrm{d}_{\mathrm{xy}}} =  \frac{\mathrm{d}_{\mathrm{xy}}}{\sigma(\mathrm{d}_{\mathrm{xy}})}$ & Significance of the transversal impact parameter $d_{xy}$ \\
            
            \bottomrule
            \end{tabular}
        \end{table}

        In order to associate the reconstruction-level jets used as an input to the models with the generator-level truth, the generator-level $\tau_{h}^{vis}$ is matched to both generator-level jets and reconstructed jets within a cone of size $\Delta R < 0.4$.
        A total of four target values are associated to each reconstruction-level jet:
        \begin{itemize}
            \item Binary label \texttt{isTau}. Set to \texttt{True} if matched to generator-level \tauh
            \item Categorical target with 6 classes corresponding to the decay modes targeted in this study
            \item Binary classification label with boolean encoding {0, 1} corresponding to the generator-level \tauh charges {-1, +1}
            \item Multi-value regression target $p^\mu$ corresponding to the generator-level visible \tauh four-momentum. 
        \end{itemize}

        In order to balance the \tauh identification task across jet kinematics, each reconstructed jet is assigned a classification weight $w_{cls}$ according to the signal and background contribution to the given bin in the polar angle--momentum ($\theta$--p) space. Weights $w_{i}$ for each bin $i$ are calculated according to \autoref{eq:weights}:

        \begin{equation}\label{eq:weights}
             w_{i} = \frac{\mathrm{min}(N_{i}^{sig}, N_{i}^{bkg})}{N_{i}^{sig}}.
        \end{equation}


        A total of 5.33 M (0.59 M) signal \z and 35.36 M (3.93 M) background $\mathrm{Z} \rightarrow qq$ jets are used for training (testing). The training dataset is further divided into train-validation subsets, allocating 20\% of the jets for validation, with the rest of them being used for training. This constitutes a total of 72-18-10\% split of the full generated dataset for training-validation-testing.

\section{Full reconstruction of \tauh}\label{sec:reconstruction}

    Hadronic tau reconstruction can be formulated as a set of supervised learning tasks that together define the properties of a \tauh candidate. In Refs.~\cite{Lange:2023gbe,TANI2025109399}, we established a unified framework addressing \tauh identification (isTau), decay mode classification (DM), and the regression of the visible transverse momentum of the generator-level \tauh (\ptvis), achieving strong performance in all tasks. In the present study, we extend this framework to a more complete reconstruction of the \tauh by covering additional properties.
    
    We now include charge reconstruction (q) as a binary classification problem and generalize the regression target to the full four-momentum ($p^{\mu}$), predicting (\ptvis, $\eta^{vis}$, $\phi^{vis}$, \massvis). The reconstruction is thus composed of four tasks:
        \begin{equation*}
            \Phi(\mathrm{jet\ features}, \mathrm{particle\ features}) \rightarrow \{\texttt{isTau}, \mathrm{DM}, \mathrm{q}, p^{\mu}\}\,\,,
        \end{equation*}
    where $\Phi$ is a trainable model.

    We consider two complementary versions of $\Phi$: SingleParTau, where each task is solved independently through per-task training, and MultiParTau, in which a single model is trained to solve all four tasks simultaneously. SingleParTau serves as a performance baseline, while MultiParTau allows the shared backbone to exploit correlations between the targets and provides a more parameter-efficient model for deployment. In all cases, the underlying model architecture is based on ParticleTransformer~\cite{Qu:2022mxj}, which operates on the reconstructed particle flow candidates forming the jet. The overview of the model layouts is shown in \autoref{fig:architecture}.


    The transformer backbone accounts for the bulk of the model size, with approximately 3.1~M parameters, and is nearly identical between SingleParTau and MultiParTau: the input embedding is shared and only the final output layer differs slightly depending on the task. An individual SingleParTau model is similar in size to the full MultiParTau model, at 3.48~M compared to 3.58~M parameters. The reduction in model complexity therefore does not arise from a smaller architecture, but from the fact that solving all four tasks with SingleParTau requires four independently trained models, amounting to a combined $4 \times 3.48~\mathrm{M} \approx 13.92$~M parameters. This is roughly four times that of MultiParTau, which instead reuses a single shared backbone across all four tasks with task-specific heads. This efficiency comes at a cost: optimizing four separate tasks through a shared network requires strict loss balancing to prevent the competing objectives from degrading overall performance.
    
    The average inference latency for a batch of 32 jets is around a few milliseconds for both SingleParTau and MultiParTau, both on a single CPU thread as well as a consumer GPU. The test machine had an Intel Core i7-10700 CPU and an Nvidia RTX 5060 Ti GPU.

    \subsection{Training strategy}
        Both the SingleParTau and MultiParTau models are trained for a maximum of 100 epochs using the AdamW optimizer~\cite{loshchilov2017decoupled} with a OneCycleLR scheduler utilizing a cosine annealing strategy. To ensure stable training and efficient convergence, we utilize 16-bit mixed-precision training. The training is done using an NVIDIA L40S GPU with a batch size of 12\,288. The input jets are zero-padded and capped at 20 constituents.


        The ParticleTransformer~\cite{Qu:2022mxj} backbone architecture used by SingleParTau and MultiParTau consists of an input embedding layer, two transformer layers, and task-specific heads. The embedding network is a feedforward architecture composed of sequential linear, normalization, and activation layers, with hidden dimensions of [256, 512, 256] in both models. The transformer layers are standard ParticleTransformer layers, using the relative $\eta$ and $\phi$ with respect to the jet axis of the constituents to define the particle attention matrix. No systematic hyperparameter or architecture optimization was done at this stage, but it is expected that a multi-objective tuning, taking into account the physics performance as well as computational performance, is straightforward and can further improve the physics reach.

        Given that the prediction targets vary both in semantic meaning and data type, each training task must be treated individually with a task-specific objective and corresponding loss function. The following paragraphs detail the training targets and loss formulations employed for the four tasks and are summarized in \autoref{tab:training-types}.

        \paragraph{\tauh identification (tagging): } formulated as a binary classification task to distinguish jets originating from \tauh decays from background qq-jets.
        Each reconstructed jet is given a tag \texttt{isTau}, set to 1 if there is a matched $\tau_{h}^{gen}$, and to 0 if there is no matched $\tau_{h}^{gen}$.
        Similarly to Ref.~\cite{Lange:2023gbe}, we use cross entropy (CE) loss for the tagging task:

        \begin{equation}\label{eq:tag-loss}
            \mathcal{L}_{\mathrm{tag}} = \mathcal{L}_{CE}(w_{i}, \theta),
        \end{equation}
        where the parameter $\theta$ denotes the model parameters, and $w_i$ the weight for the individual jet $i$ based on its kinematics.

        \paragraph{Kinematic reconstruction: } a multi-target regression task aimed at reconstructing the full visible \tauh four-momentum. We chose the ($p_T, \eta, \varphi, m$) parametrization of the four-momentum to be reconstructed, but other parametrizations are possible. We have chosen mass $m$ as the fourth component as energy E is largely redundant given $m << p_T \cdot cosh(\eta)$.

        Each target contributes to the overall kinematics loss as:

        \begin{align}
            \Lambda_{p} &= \log\left(\frac{p_T^{\text{gen-}\tau}}{p_T^{\text{reco-jet}}}\right) \\
            \Lambda_{\eta} &= \Delta^{\pm}\eta, \\
            \Lambda_{s\varphi} = \sin\left(\Delta^{\pm}\varphi\right)&,\, \Lambda_{c\varphi} = \cos\left(\Delta^{\pm}\varphi\right) \\
            \Lambda_{m} &= \log\left(\frac{m^{\text{gen-}\tau}}{m^{\text{reco-jet}}}\right).
        \end{align}
        Here the $\Delta^{\pm}$ denotes the signed difference of the given property for the generator-level visible tau and the reconstruction-level jet. In the case of $\varphi$ we use the wrapped difference~\cite{cremers2018circular} as shown in \autoref{eq:dphi}:
        
        \begin{equation}\label{eq:dphi}
            \Delta^{\pm}\varphi = \mathrm{atan2}\bigl(\sin(\Delta^{\pm}\varphi),\, \cos(\Delta^{\pm}\varphi)\bigr)
        \end{equation}
        
        The first component of the overall kinematics loss is ``chord'' loss, which is the hypotenuse of a triangle with the sides corresponding to $\Lambda_{s\varphi}$ and $\Lambda_{c\varphi}$. All other components of the kinematics loss use Huber loss~\cite{10.1214/aoms/1177703732} with the delta parameter set to 1.0 as shown in \autoref{eq:kin-loss}:
        \begin{equation}\label{eq:kin-loss}
            \mathcal{L}_{\text{kin}} = \frac{\lambda_{chord} \mathcal{L}_{chord}(\Lambda_{s\varphi}, \Lambda_{c\varphi}) + \sum_{i} \lambda_{i} \mathcal{L}_{\text{Huber}}\bigl(\Lambda_i\bigr)}{\lambda_{chord} + \sum_{i} \lambda_i}
        \end{equation}
        with $\mathrm{i} \in \{\mathrm{p}, \eta, \mathrm{m}\}$. As the \tauh visible mass is noisy due to the neutrinos in the \tauh decay escaping detection, we choose the weight for the mass component of the kinematics loss $\lambda_{m} = 0.2$ and set all other $\lambda_i$ to 1, in order to down-weight the mass term's contribution to the kinematics loss.

        \paragraph{Charge reconstruction:} formulated as a binary classification problem, assigning each \tauh candidate to one of the two possible charges, $\mathrm{q}^{\mathrm{true}}=\pm1$. For training purposes, the physical charge labels are mapped to binary class indices according to
        \begin{equation}
            y^{\mathrm{charge}} =
            \begin{cases}
                0, & \mathrm{q}^{\mathrm{true}}=-1 \\
                1, & \mathrm{q}^{\mathrm{true}}=+1
            \end{cases}.
        \end{equation}
        
        The network is designed with a two node output layer, where each node produces a logit corresponding to one of the two charge hypotheses. The model is trained using the cross-entropy loss function $\mathcal{L}_{charge} = \mathcal{L}_{CE}$.
        During inference, the predicted class index is mapped back to the corresponding physical charge assignment, $\mathrm{q}^{\mathrm{pred}}=\pm1$.

        \paragraph{Decay mode classification:}
        a task assigning each \tauh candidate to one of six categories derived from the \gls{hps}~\cite{cms2012performance, cms2016reconstruction} decay mode indexing scheme, corresponding to the dominant \tauh decay modes defined by the multiplicity of charged hadrons ($h^{\pm}$) and neutral pions ($\pi^0$). The target class index $y_i^{\mathrm{DM}} \in \{0,\dots,5\}$ is defined as:
        \begin{equation}
            y_i^{\mathrm{DM}} =
            \begin{cases}
                0, & h^{\pm} \\
                1, & h^{\pm}\pi^0 \\
                2, & h^{\pm} + \geq 2\pi^0 \\
                3, & h^{\pm}h^{\mp}h^{\pm} \\
                4, & h^{\pm}h^{\mp}h^{\pm} + \geq \pi^0 \\
                5, & \text{rare modes}
            \end{cases}.
        \end{equation}
        The target labels are one-hot encoded and the model is trained using the standard cross-entropy loss for multi-class classification $\mathcal{L}_{DM} = \mathcal{L}_{CE}$.

        \begin{table}[H]
        \centering
        \caption{Training types and loss functions for the four tasks needed for \tauh reconstruction.}
        \label{tab:training-types}
            \begin{tabular}{lll}
            \toprule

            \textbf{Task} & \textbf{Training type} & \textbf{Loss function} \\
            \midrule
            \tauh identification & binary classification & \gls{ce} loss \\
            $p^\mu$ reconstruction & multi-target regression & Huber loss + ``chord'' loss \\
            decay mode classification & multi-class classification & cross entropy (\gls{ce}) loss \\
            charge identification & binary classification &cross entropy (\gls{ce}) loss \\
            
            \bottomrule
            \end{tabular}
        \end{table}

        \subsection{Unified model for full \tauh reconstruction}

        Training the multi-task model requires careful balancing of the loss across the task-specific heads. The difficulty arises from potential gradient interference, where updates for one task may negatively impact the performance of others. To address this, the multi-task MultiParTau model employs the PCGrad algorithm~\cite{yu2020gradient} that dynamically modifies the gradients of task-specific losses by reducing gradient interference between the four tasks during training. This helps the backbone model to learn features beneficial to all objectives aiming to reduce the negative weight updates on any one task in the process. As a safeguard, the gradients are clipped with a max norm of 1.0 after PCGrad merging and before the optimizer step to avoid destabilizing gradients.

\begin{figure}[H]
            \centering
            \includegraphics[width=\textwidth]{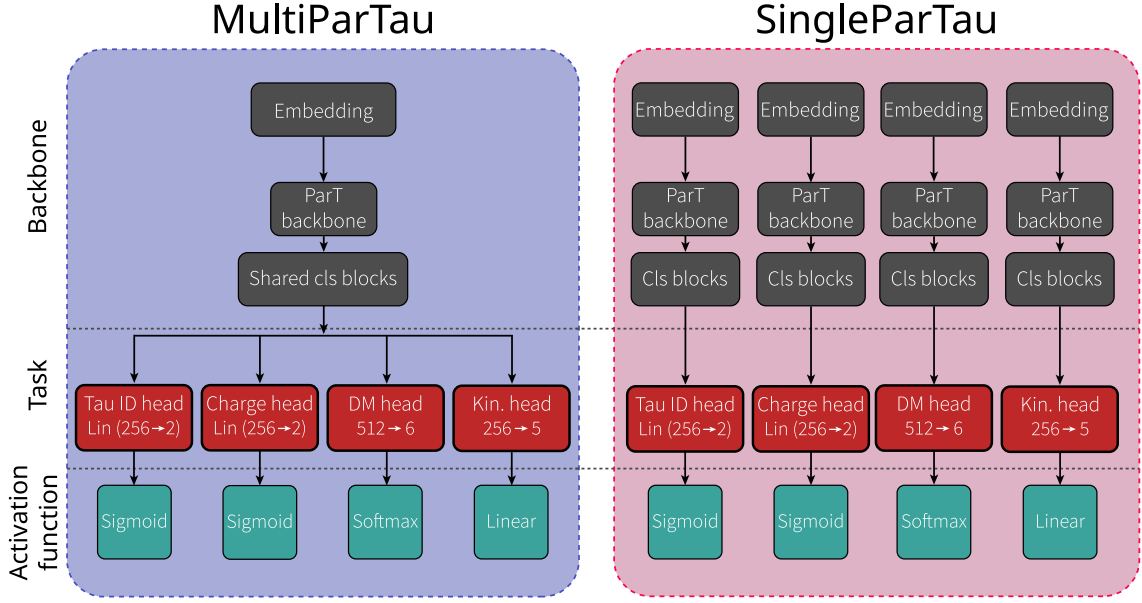}
            \caption{Schematic of the model architectures of the two training approaches: (\textbf{Left}) The multi-task model features a common backbone shared among all four tasks.  (\textbf{Right}) The dedicated model has a separate backbone trained for each task. All task-specific heads consist of two linear layers, with the second layer having an output dimension of 128. Given the increased difficulty of the kinematic regression task, the kinematic head is the only one for which dropout is not applied.}
            \label{fig:architecture}
        \end{figure}

        As shown in \autoref{fig:architecture}, the MultiParTau model consists of a shared embedding layer and a common backbone that processes the constituents of the jet. Task separation is achieved using four dedicated CLS tokens that attend to the shared backbone output. Each task-specific token then passes through its own readout head to produce the final predictions. This architecture allows the model to learn a unified representation of the jet while maintaining specialized branches for each reconstruction task.

        The total training loss for MultiParTau is a weighted combination of the four task-specific losses:
        \begin{equation}
            \mathcal{L}_{\text{multi}} = \sum_{i} \lambda_i w_i \mathcal{L}_{i}
            \quad \text{with } i \in \{\text{tag},\,\text{kin},\,\text{DM},\,\text{charge}\},
        \end{equation}
        where $w_i$ denotes the kinematic-based sample weight.

        \subsection{Training convergence}
\begin{figure}[H]
            \centering
            \includegraphics[width=0.85\textwidth]{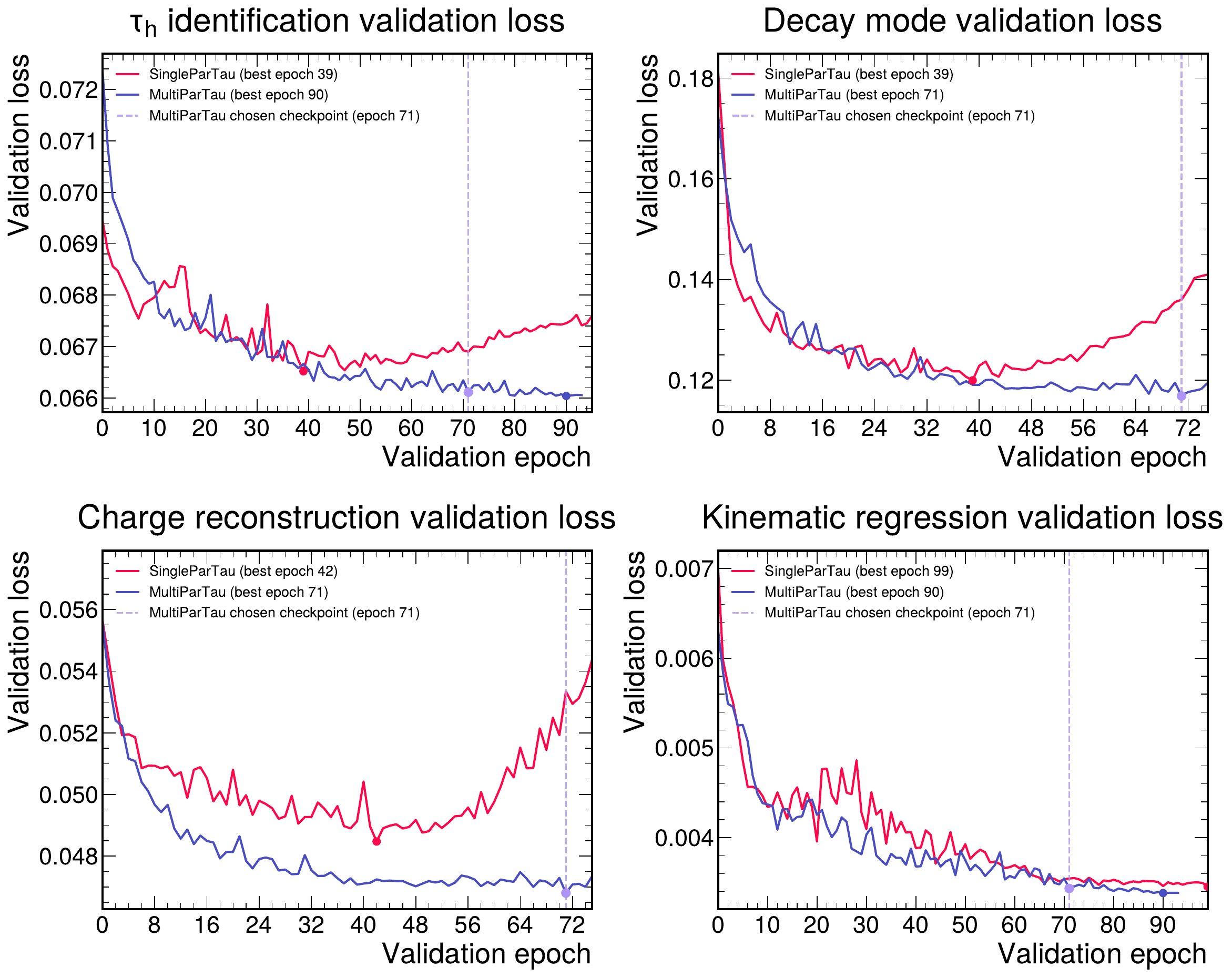}
            \caption{Evolution of the validation loss during training for the \tauh identification, decay mode classification, charge reconstruction, and kinematic regression tasks. Absolute validation loss values are shown for SingleParTau and MultiParTau, with the best validation epoch indicated. For MultiParTau, the model checkpoint selected for evaluation is also shown.}
            \label{fig:losses}
        \end{figure}

        The convergence behaviour of the models is shown in \autoref{fig:losses}. The training is stable and converges for both approaches across all four reconstruction tasks. MultiParTau reaches lower validation losses and shows smooth convergence throughout training, while SingleParTau exhibits a gradual increase in validation loss at later epochs for the classification tasks, indicating mild overfitting. For SingleParTau, the model checkpoint with the lowest validation loss is selected independently for each task and used for the performance evaluation. For MultiParTau, the final model checkpoint is selected according to the total validation loss summed across all four tasks. The vertical dashed line in \autoref{fig:losses} indicates the chosen checkpoint, which represents the best overall compromise among the four reconstruction objectives.
        
\section{Results}\label{sec:results}

The performance of the hadronic tau reconstruction models is characterized using dedicated, detailed metrics for each reconstruction task.
    
\begin{figure}[H]
            \centering
            \includegraphics[width=0.48\textwidth]{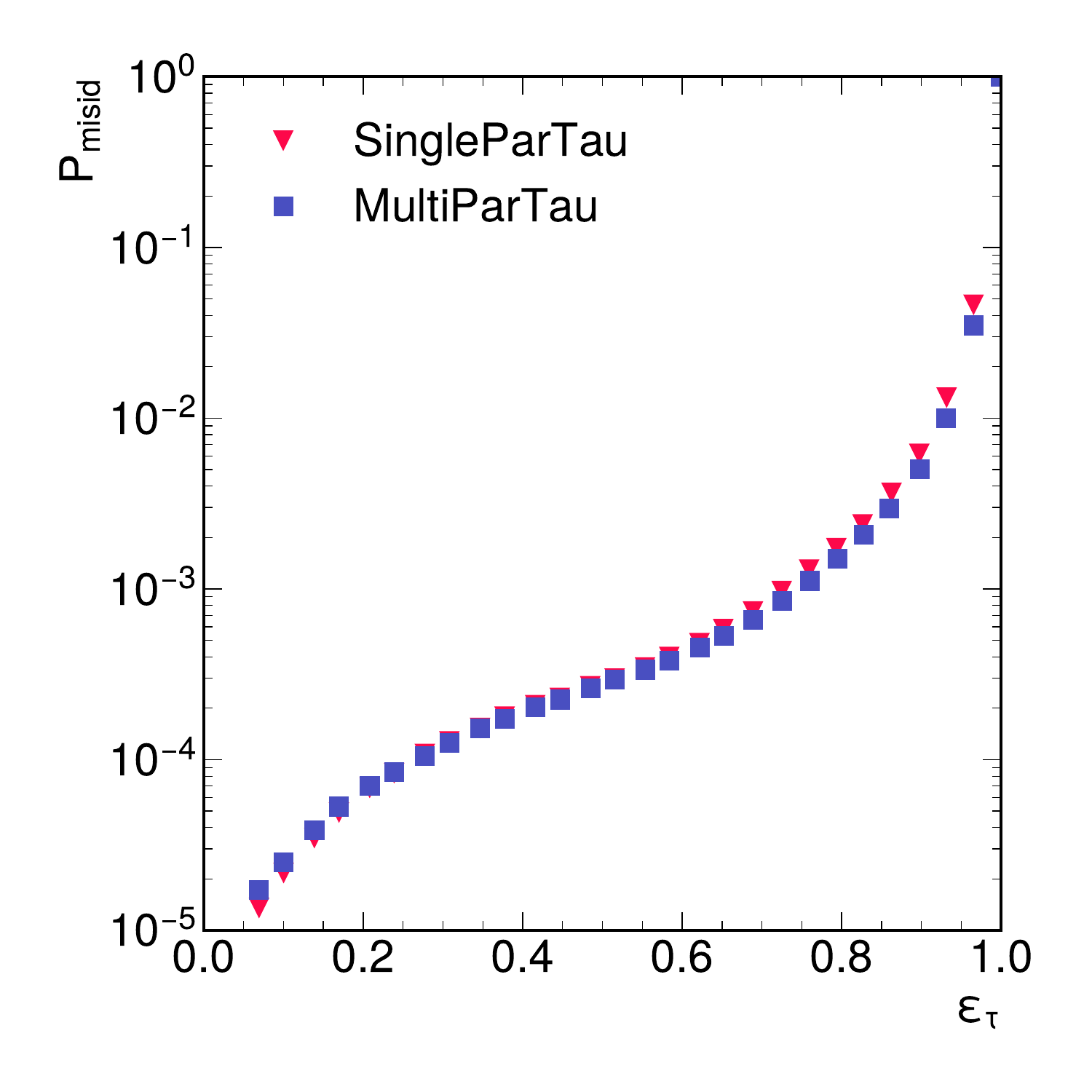}
            \includegraphics[width=0.458\textwidth]{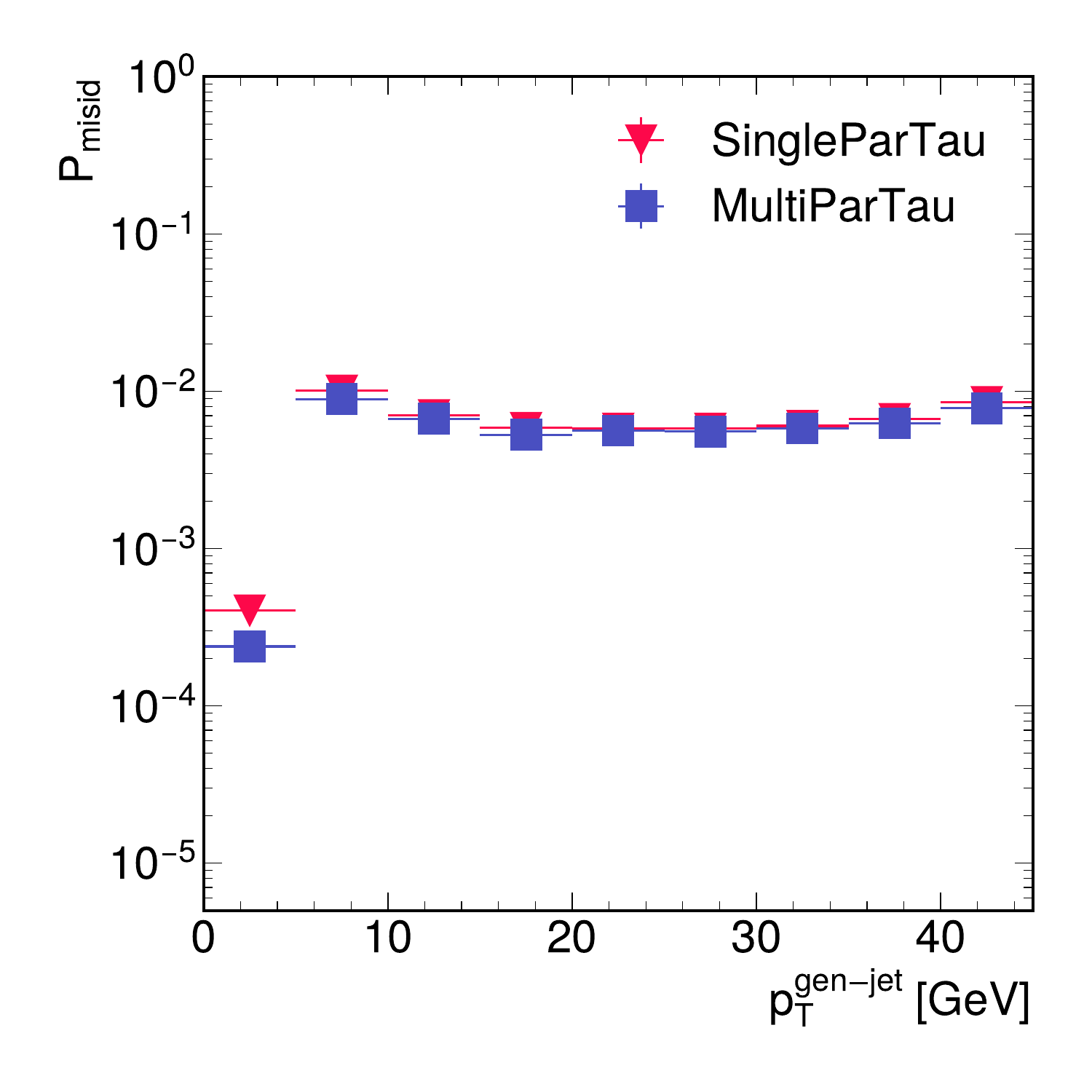}
            \caption{
            (\textbf{Left}) The jet mis-identification rate $P_{\mathrm{misid}}$ as a function of the \tauh identification efficiency $\epsilon_\tau$ is nearly identical for SingleParTau and MultiParTau. (\textbf{Right}) The jet mis-identification rate per generator-level background jet $p_T$ bin at a global average \tauh identification efficiency of $80\%$. The fake rate is $\mathcal{O}(10^{-4})$ in the 0--5~GeV bin, rising to $\mathcal{O}(10^{-2})$ above 5 GeV, with both models performing comparably across the full $p_T$ range.
            }
            \label{fig:results_tauid}
      \end{figure}

The \textbf{tau identification} performance is evaluated using the receiver operating characteristic (ROC) curve, shown in \autoref{fig:results_tauid} (left), which displays the jet mis-identification rate $P_{\mathrm{misid}}$ as a function of the \tauh identification efficiency $\epsilon_{\tau}$. 
The signal efficiency is evaluated on $Z/\gamma^* \rightarrow \tau\tau$ jets matched to generator-level $\tau_h$, and the mis-identification rate on $Z/\gamma^* \rightarrow qq$ background jets, both required to satisfy $10^\circ < \theta < 170^\circ$ at generator and reconstruction level, with no minimum $p_T$ requirement, in order to include low-$p_T$ \tauh candidates relevant for the TeraZ physics program. The two approaches exhibit nearly identical performance over the full range of operating points, with MultiParTau performing slightly better.
\autoref{fig:results_tauid} (right) shows the jet mis-identification rate per generator-level background-jet $p_T$ bin at a fixed average \tauh identification efficiency of $80\%$. The mis-identification rate is significantly lower in the 0--5~GeV bin, where it is $\mathcal{O}(10^{-4})$, rising to $\mathcal{O}(10^{-2})$ for jets above 5 GeV. This is because the score threshold is set for a global average efficiency of $80\%$, which corresponds to a signal efficiency of only $\sim40$--45$\%$ 
in the 0--5~GeV bin, resulting in an effectively tighter selection on background jets in that region.

\begin{figure}[H]
            \centering
            \includegraphics[width=0.48\textwidth]{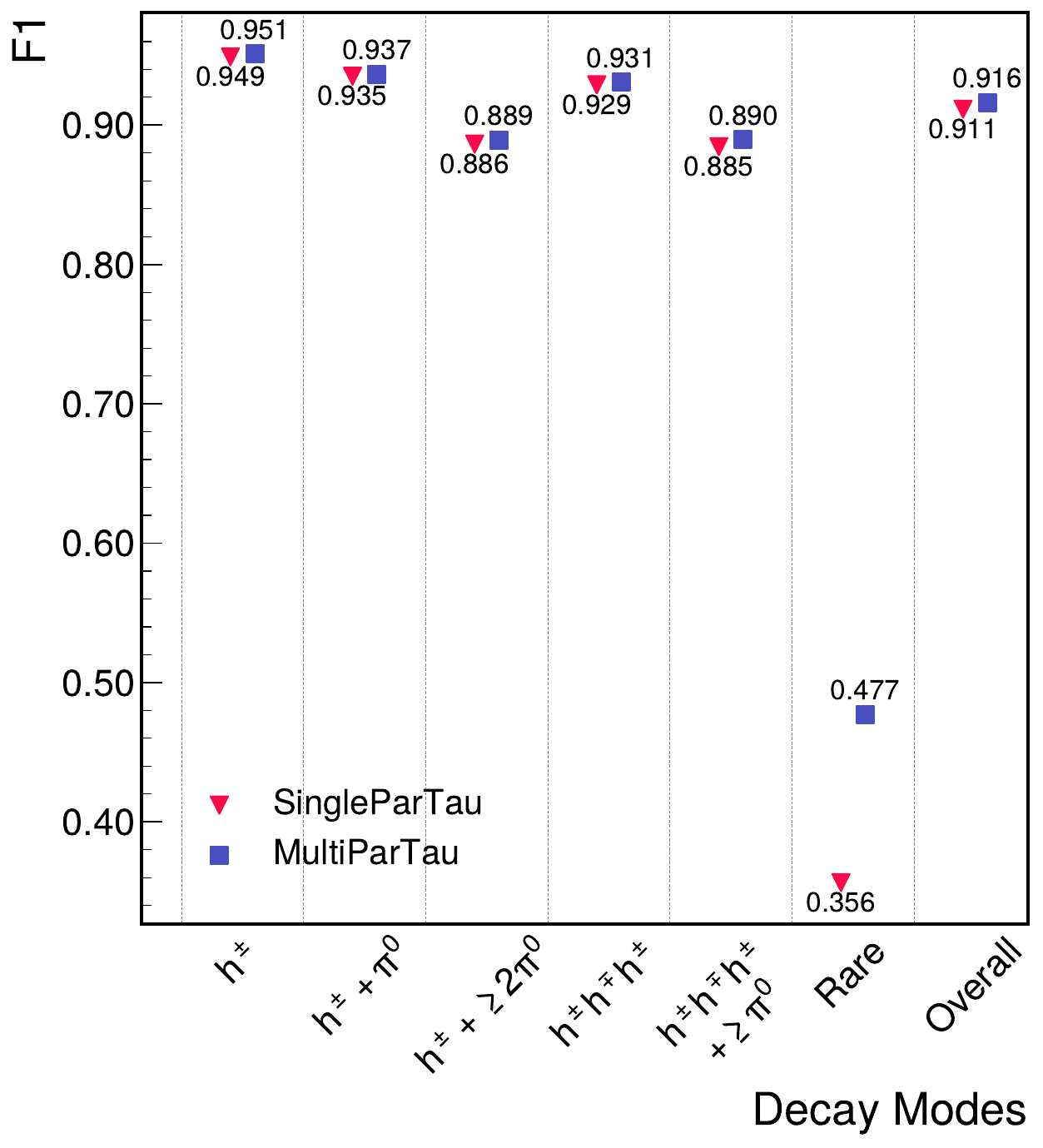}
            \includegraphics[width=0.51\textwidth]{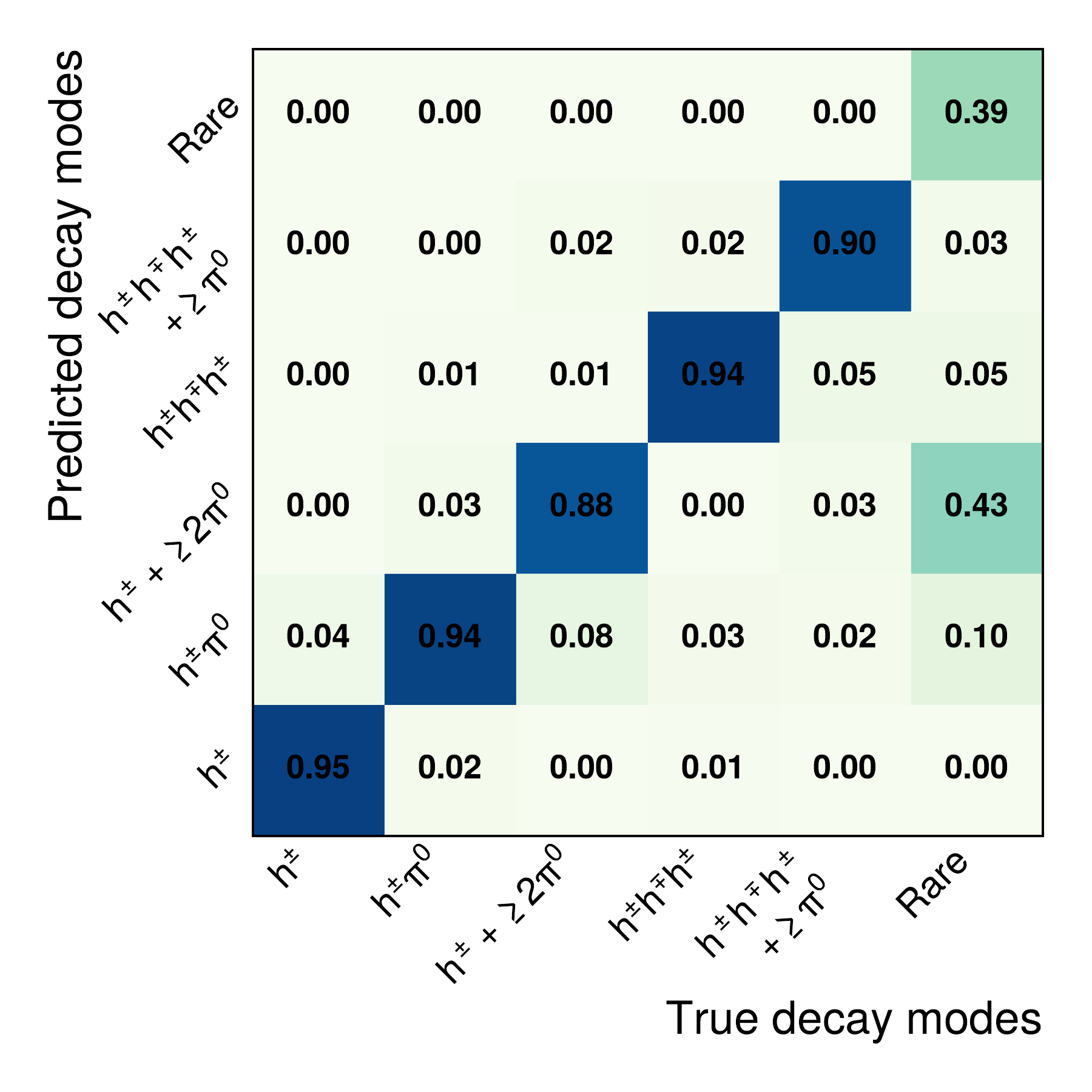}
            \caption{(\textbf{Left}) While both MultiParTau and SingleParTau have a similar performance, the unified model consistently has a slightly better performance measured in terms of the F1 score, the harmonic mean of precision and recall. (\textbf{Right}) \tauh class-wise classification performance normalized over generator-level \tauh decay modes for MultiParTau.}
            \label{fig:results_DM}
      \end{figure}

For the \textbf{decay mode classification}, it is important to note that the different \tauh decay modes are not equally populated in the dataset. While the dominant channels are well represented, categories with multiple neutral pions as well as rare decays contain significantly fewer samples. To evaluate the classification performance across all decay modes, we use the class-wise F1 score, which combines the precision and recall of the classifier into a single metric to measure how accurately each decay mode is identified. \autoref{fig:results_DM} (left) compares the F1 score of the SingleParTau and MultiParTau models for the individual generator-level \tauh decay modes, where the ``Overall'' category represents the average performance across all channels, weighted by the branching fractions of the corresponding \tauh decays.

Both approaches achieve very similar F1 scores of approximately $0.89$--$0.95$ for the dominant channels. The best performance is obtained for the single charged hadron ($h^\pm$) decay mode, while the classification becomes more challenging for channels containing multiple neutral pions as well as for the statistically limited rare category. The MultiParTau model achieves slightly higher F1 scores across all decay modes, although the difference between the two approaches remains small. \autoref{fig:results_DM} (right) shows the confusion matrix for the MultiParTau model, normalized to the generator-level true decay modes. The strong diagonal structure indicates that the classifier correctly identifies the true channel in the majority of events, with correctly reconstructed fractions reaching approximately $90$--$95\%$ for the dominant channels. The largest off-diagonal contributions arise from migrations between categories with similar final state signatures, in particular between decay modes differing only in the number of reconstructed neutral pions. Mis-classifications involving the rare category are also observed more frequently due to the limited number of available events in these channels.

In order to evaluate the performance on the \textbf{kinematic reconstruction} task, we use the median $\Delta R$ between the predicted and generator-level \tauh candidates for the angular component. For the momentum, we determine the resolution of the reconstructed visible transverse momentum \ptvis as a function of the generator-level visible \tauh $p_T$. The \ptvis resolution is quantified using the interquartile range (IQR) of the $p_T^{\mathrm{pred}} / p_T^{\mathrm{gen}}$ distribution, normalized to its median, providing a robust measure of the distribution width where smaller values correspond to a better momentum resolution while reducing sensitivity to outliers. In addition, two-dimensional distributions of the predicted and generator-level kinematic quantities are used to illustrate the agreement and resolution of the reconstructed $p^\mu$ components.

\begin{figure}[H]
            \centering
            \includegraphics[width=0.48\textwidth]{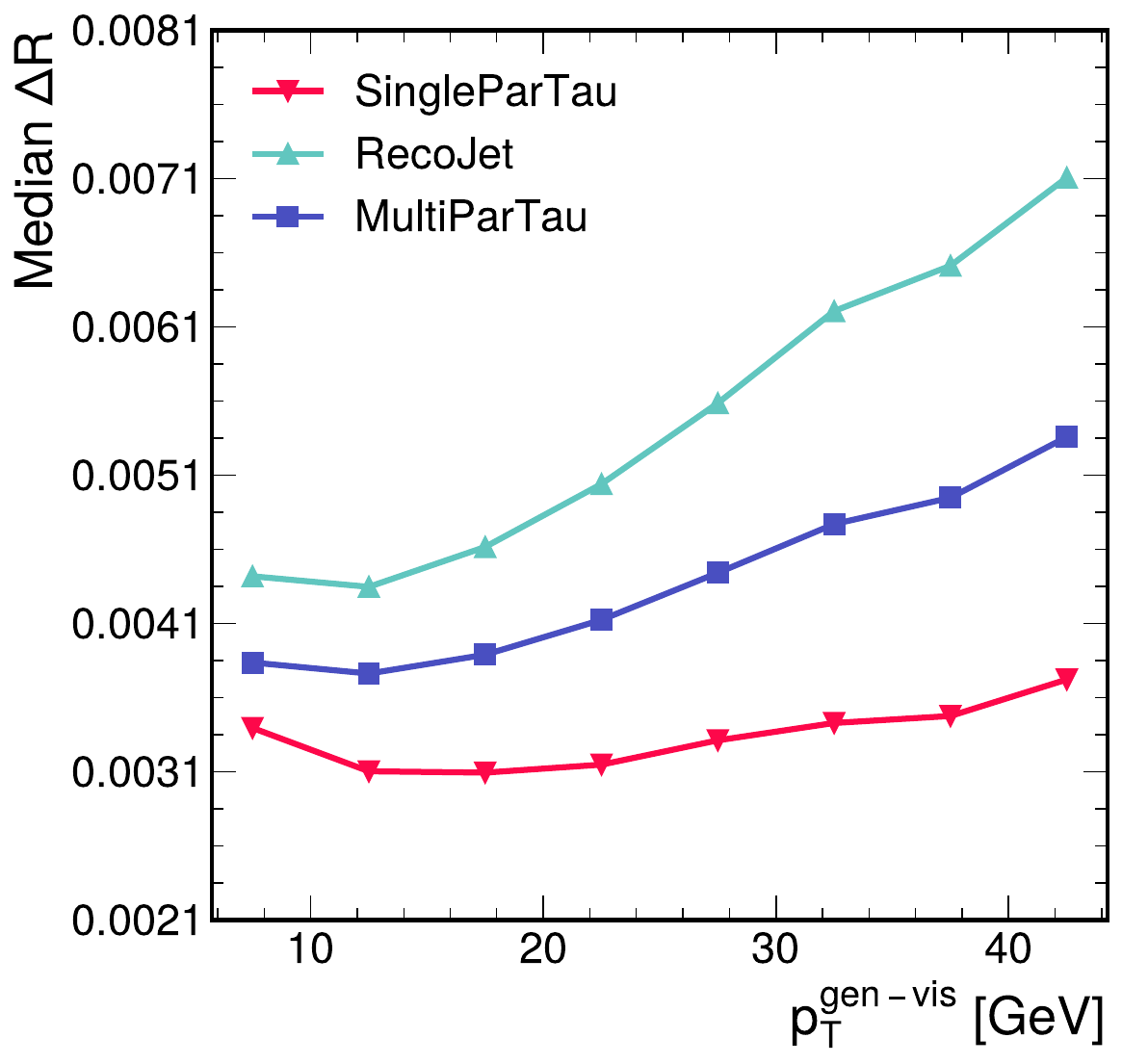}
            \includegraphics[width=0.48\textwidth]{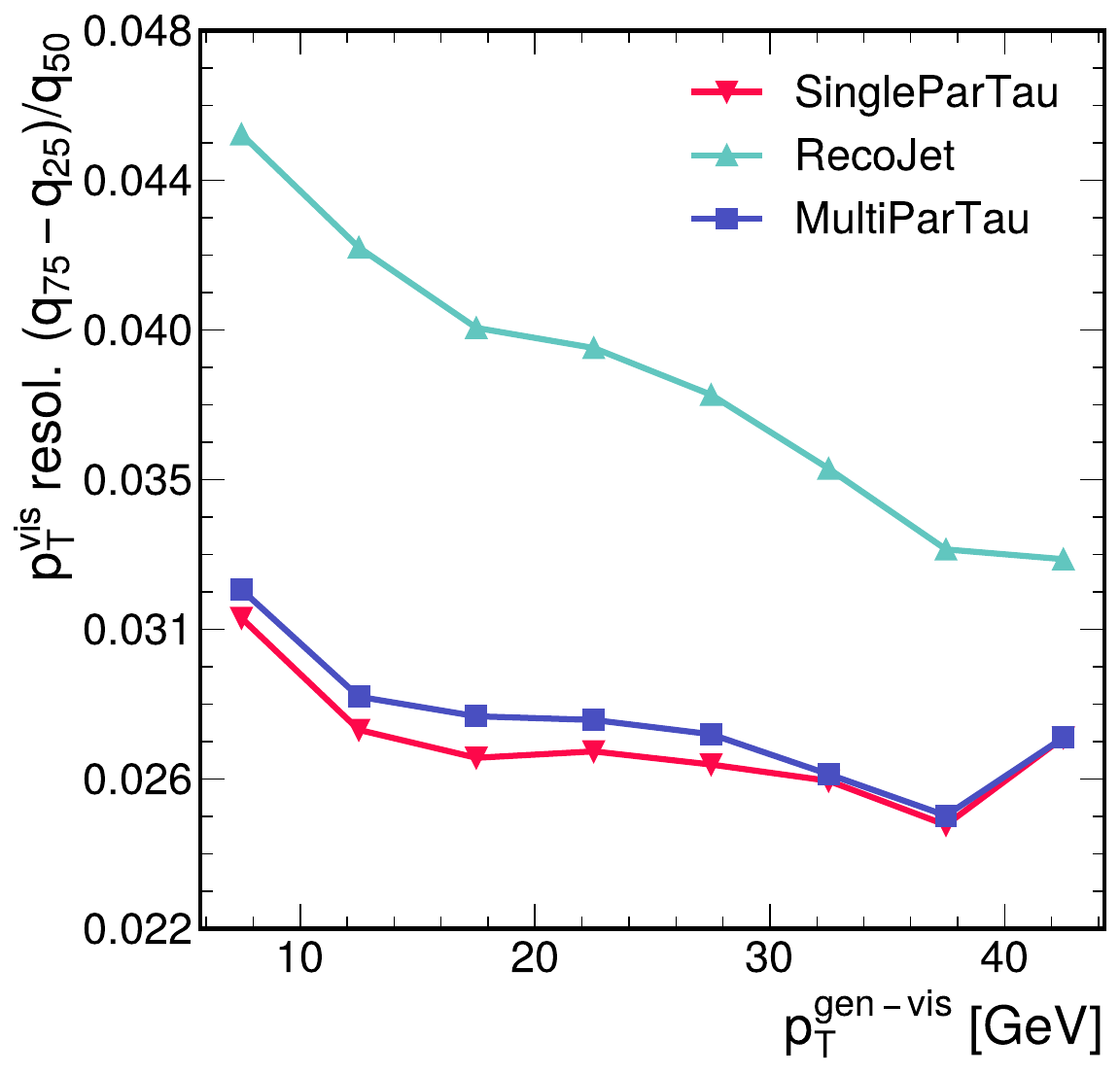}
            \caption{The dependence of the reconstructed \tauh $\mathrm{median}(\Delta R)$ (\textbf{left}) and  \ptvis (\textbf{right}) on the generator-level visible \tauh $p_T$.}
            \label{fig:results_regression}
        \end{figure}

\autoref{fig:results_regression} (left) shows the median $\Delta R$ between the predicted and generator-level \tauh candidates. The SingleParTau model achieves the best angular resolution over the full $p_T$ range, with median $\Delta R$ values of approximately $3.1$--$3.7\times10^{-3}$. The MultiParTau model exhibits comparable angular performance, with median $\Delta R$ values of approximately $3.8$--$5.4\times10^{-3}$.

 As a reconstruction-level cross-check, the same quantities are also computed using the original reconstructed jet features, denoted as RecoJet. This approach yields significantly larger median $\Delta R$ values of approximately $4.3$--$7.1\times10^{-3}$, indicating improved angular resolution for the ML-based approaches. All methods exhibit a gradual degradation in angular resolution with increasing \tauh transverse momentum. \autoref{fig:results_regression} (right) shows the corresponding \ptvis resolution. The SingleParTau model again provides the best overall performance, achieving an IQR-based momentum resolution of approximately $2.51$--$3.11\%$ over the studied $p_T$ range. The MultiParTau model exhibits nearly identical performance, achieving resolutions of approximately $2.54$--$3.20\%$, compared to $3.28$--$4.53\%$ for the RecoJet approach. The momentum resolution improves with increasing $p_T$ up to approximately 40 GeV, after which a slight deterioration is observed, reflecting the reduced number of available high-$p_T$ jets in the dataset.

\begin{figure}[H]
        \centering
    
        \begin{subfigure}[b]{0.48\textwidth}
            \centering
            \includegraphics[width=\textwidth]{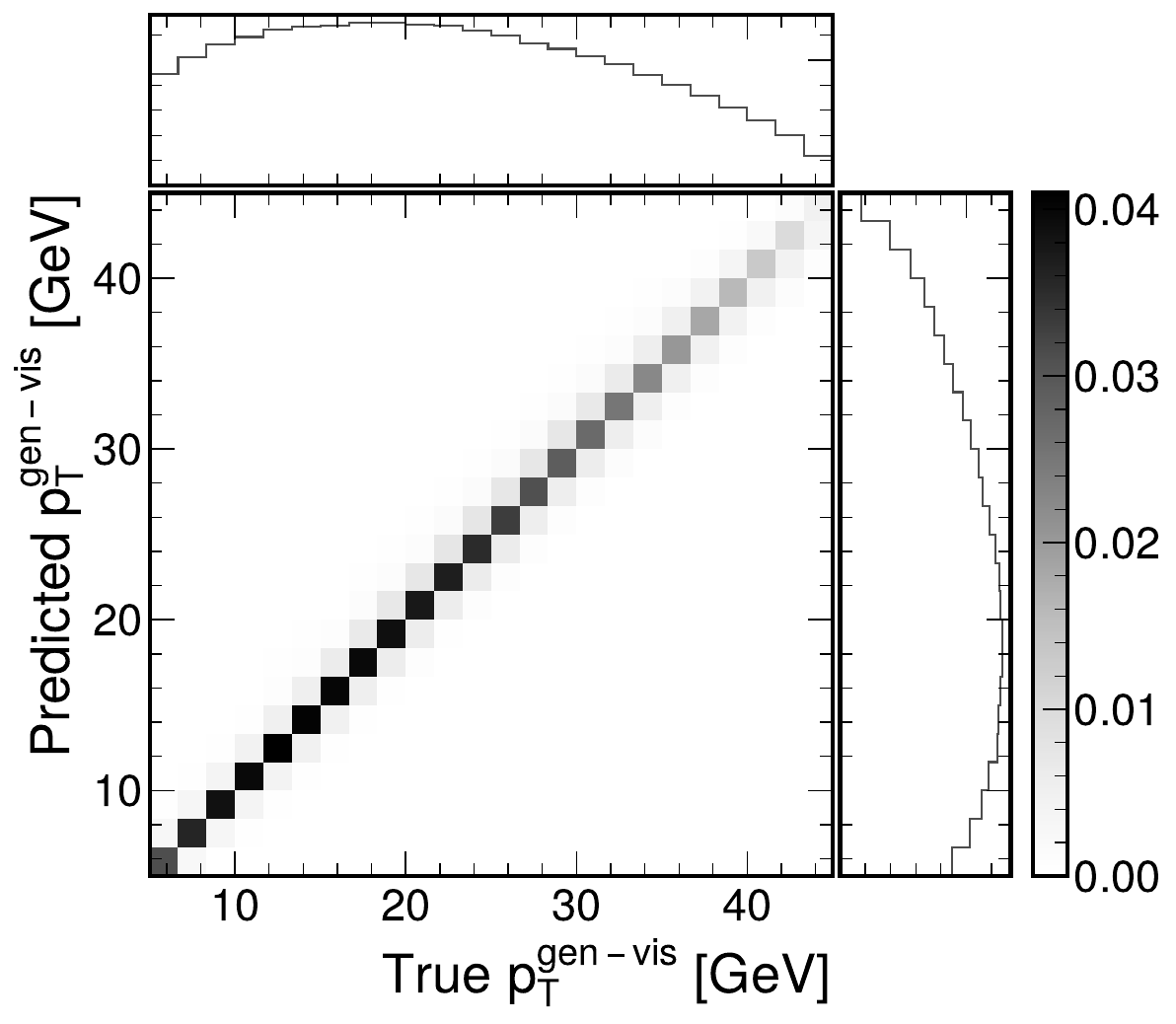}
            \label{fig:sub1}
        \end{subfigure}
        \hfill
        \begin{subfigure}[b]{0.48\textwidth}
            \centering
            \includegraphics[width=\textwidth]{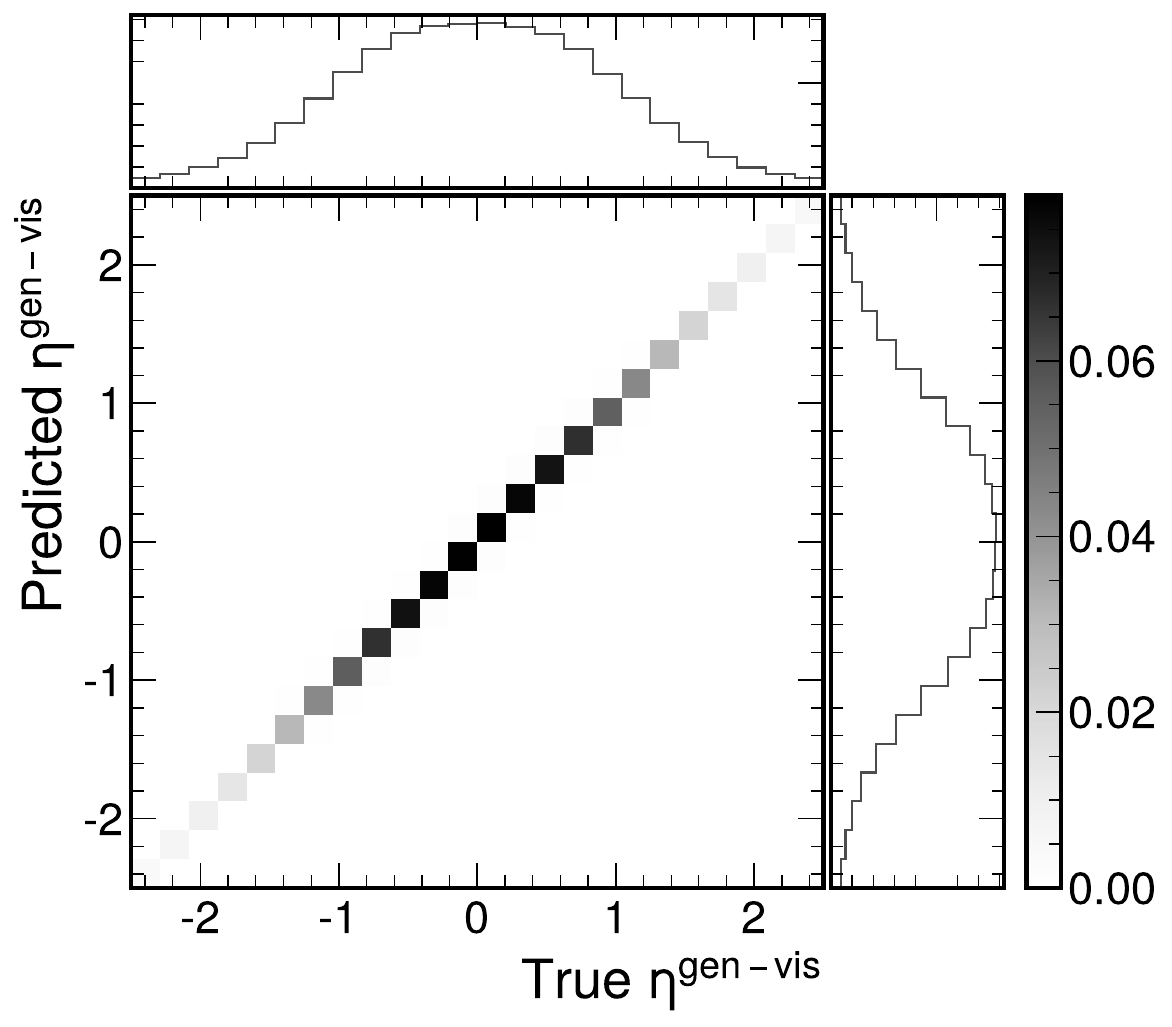}
            \label{fig:sub2}
        \end{subfigure}
    
        \vspace{0.5cm}
    
        \begin{subfigure}[b]{0.48\textwidth}
            \centering
            \includegraphics[width=\textwidth]{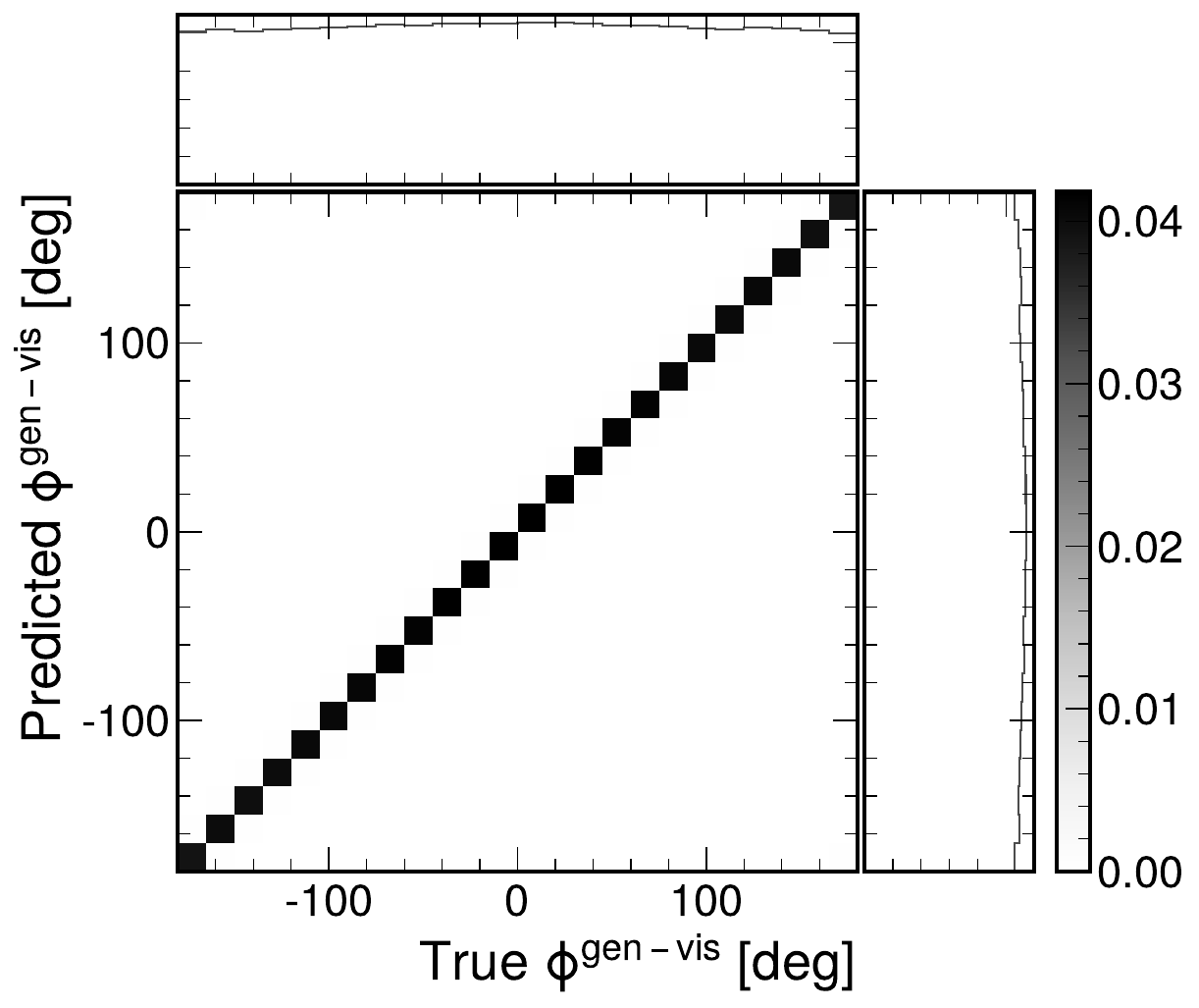}
            \label{fig:sub3}
        \end{subfigure}
        \hfill
        \begin{subfigure}[b]{0.48\textwidth}
            \centering
            \includegraphics[width=\textwidth]{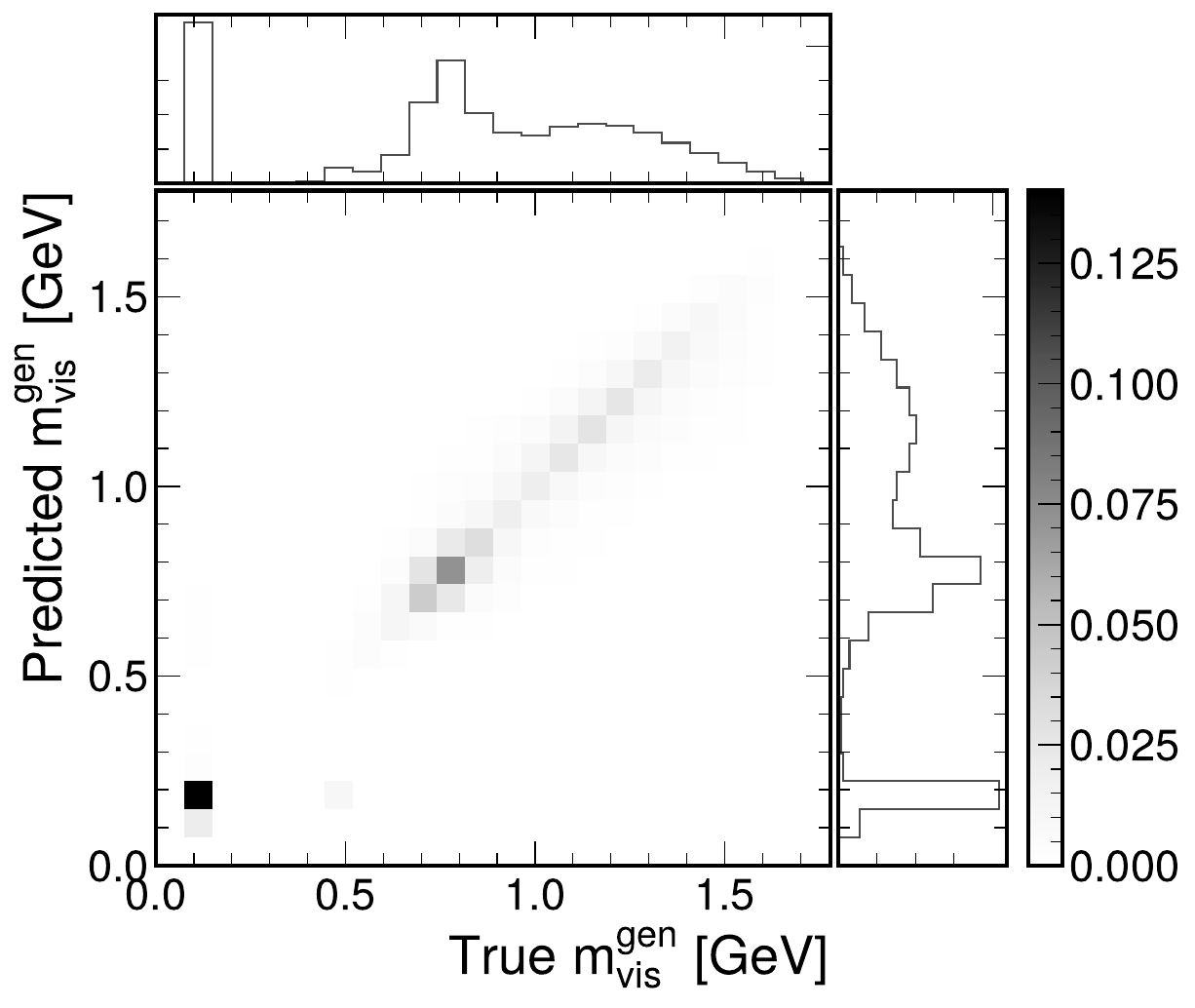}
            \label{fig:sub4}
        \end{subfigure}
    
        \caption{True and predicted $p^\mu$ components for SingleParTau.}
        \label{fig:main}
    \end{figure}

\autoref{fig:main} shows how well the four $p^\mu$ components are reconstructed for the SingleParTau model. Each panel shows a two-dimensional histogram  of the predicted value against the generator-level truth, where a perfect reconstruction would place all jets exactly on the diagonal. For the two angular variables, the jet density sits tightly along the diagonal across the full range, consistent with the small median $\Delta R$ values seen in \autoref{fig:results_regression} (left). The visible transverse momentum is also well reconstructed, with the jet density following the diagonal closely. At higher $p_T$ the spread grows, due to the limited number of high-$p_T$ jets in the dataset. The visible invariant mass is noticeably harder to reconstruct, with the distributions spread more broadly and clustering near a few preferred values. This is because \massvis varies strongly with the \tauh decay mode, with each decay mode populating a different mass region, but the kinematic head receives no decay mode information and must infer the correct mass purely from the particle flow candidates of the reconstructed jet. Furthermore, since the \tauh decay produces a neutrino that escapes detection, \massvis does not take a single fixed value, which further complicates the regression.

The \textbf{charge reconstruction} performance is evaluated using the charge mis-identification rate,
defined as the fraction of matched generator-level \tauh candidates for which the predicted charge differs from the true charge.
As a reference, we compare the ML models to a conventional jet charge observable~\cite{CMS:2016yuu} denoted as QKappa:
    \begin{equation}
        Q^\kappa = \frac{1}{\left(p_T^{\mathrm{jet}}\right)^\kappa}\sum_i q_i \left(p^{i}_{T}\right)^\kappa \,,
    \end{equation}
where the sum is taken over all particle flow candidates in the jet, with $q_i$ and $p_T^i$ denoting the charge and transverse momentum of the $i$-th candidate. The parameter $\kappa$ controls the relative weighting of low- and high-$p_T$ particle flow candidates. This observable provides a simple physics motivated estimate of the \tauh charge based on the momentum weighted distribution of the particle flow candidate charges.

\begin{figure}[H]
        \centering
        \includegraphics[width=0.48\textwidth]{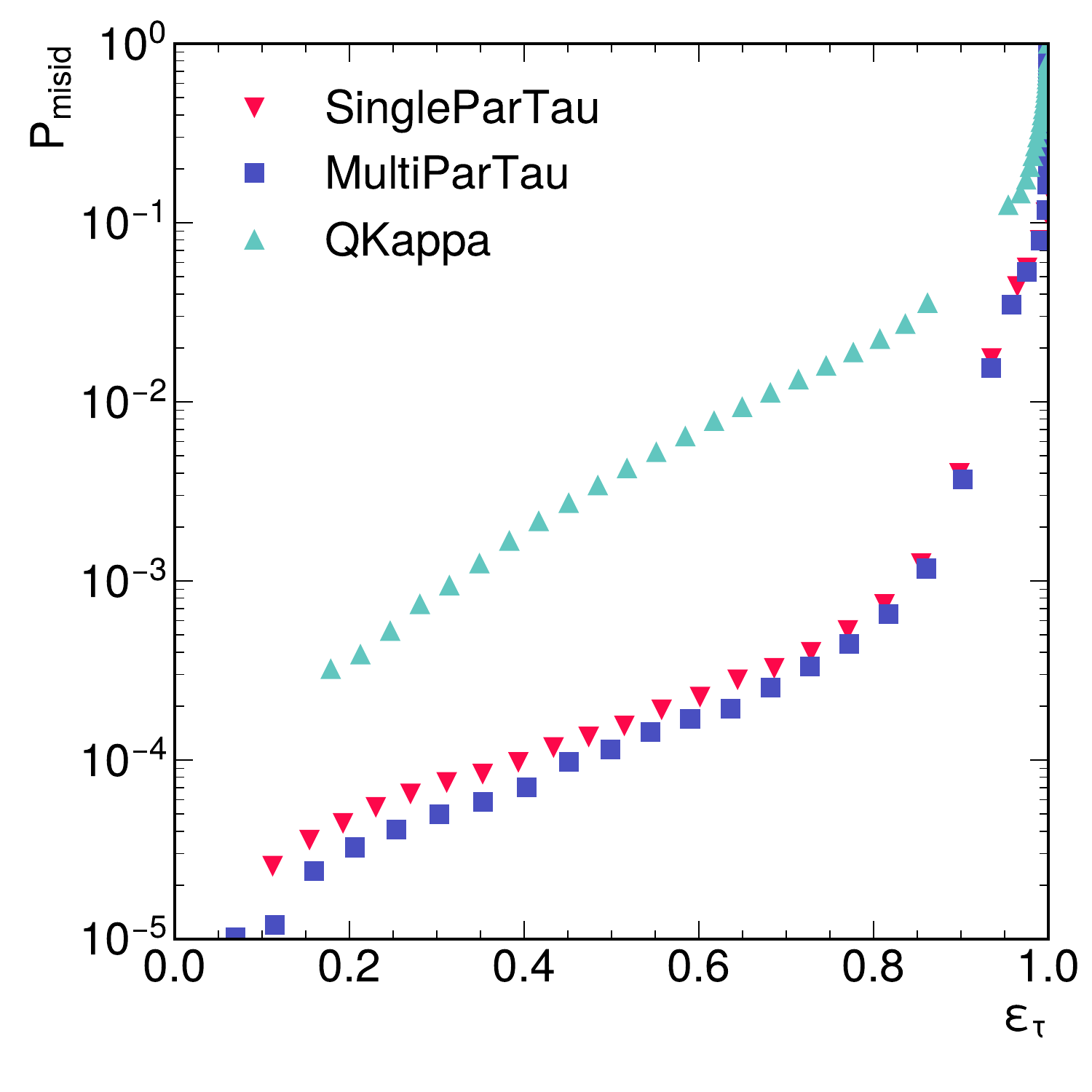}
        \includegraphics[width=0.48\textwidth]{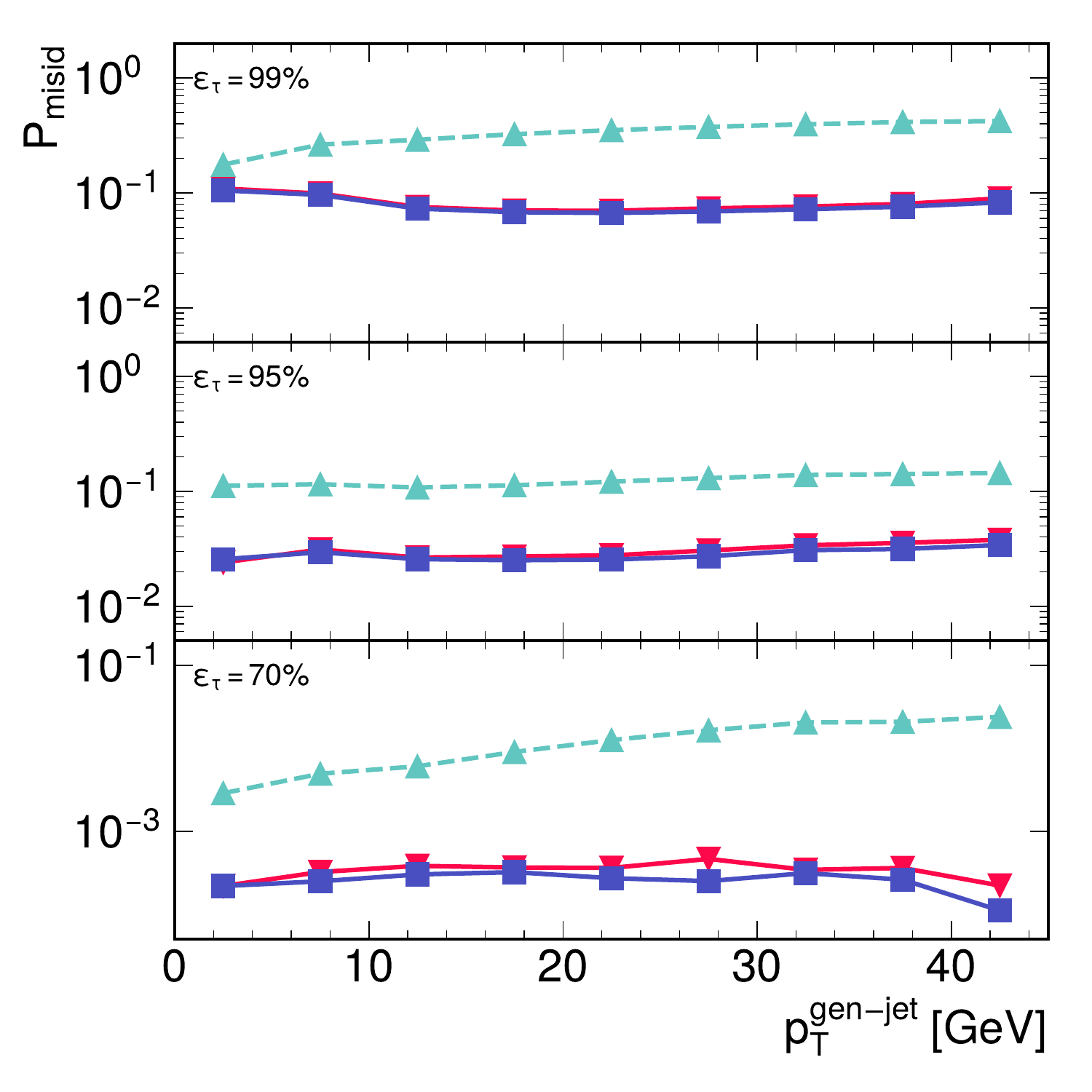}
        \caption{  Charge reconstruction performance for SingleParTau, MultiParTau, and the QKappa baseline. (\textbf{Left}) Charge mis-identification rate $P_{\mathrm{misid}}$ as a function of the charge identification efficiency $\varepsilon_{\tau}$. (\textbf{Right}) $P_{\mathrm{misid}}$ as a function of the generator-level visible \tauh transverse momentum at fixed charge identification efficiencies of 99\%, 95\%, and 70\%. Both models show very similar performance and significantly outperform the QKappa baseline.}
        \label{fig:results_charge}
    \end{figure}

\autoref{fig:results_charge} (left) shows the charge mis-identification rate $P_{\mathrm{misid}}$ as a function of the charge identification efficiency $\varepsilon_{\tau}$. The ML-based approaches exhibit very similar performance over the full efficiency range. Up to 80\% efficiency, both SingleParTau and MultiParTau achieve charge mis-identification rates below $10^{-3}$, with only small differences between them. The QKappa baseline shows a mis-identification rate roughly one to two orders of magnitude larger than the ML-based approaches, demonstrating that the models exploit the particle flow candidate content of the reconstructed jet considerably more effectively than a fixed momentum weighted jet charge estimator. \autoref{fig:results_charge} (right) shows $P_{\mathrm{misid}}$ as a function of the generator-level visible \tauh transverse momentum at fixed charge identification efficiencies of 99\%, 95\%, and 70\%. At all three working points, both ML-based models achieve a stable mis-identification rate across the full $p_T$ range, with no significant degradation at low transverse momentum. The QKappa baseline exhibits a substantially higher and more $p_T$-dependent mis-identification rate, particularly at high-$p_T$ where the momentum weighting of the jet charge estimator loses discriminating power. Both SingleParTau and MultiParTau perform comparably across all working points and $p_T$ values, with MultiParTau maintaining a slight advantage consistent with the behaviour observed in the other classification tasks.

\section{Summary and Outlook}\label{sec:outlook}

    In this paper, we extend the unified ParticleTransformer-based \tauh reconstruction approach established in Ref.~\cite{TANI2025109399} to a more complete set of reconstruction tasks. In addition to $\tau_h$ identification and decay mode classification, we incorporate charge identification ($\mathrm{q}$) and full visible four-momentum regression ($p^\mu = (\ptvis, \eta^{vis}, \phi^{vis}, \massvis)$), bringing the total to four addressed tasks. We also introduce two complementary training strategies: SingleParTau, where a dedicated model is trained independently for each task, and MultiParTau, where a single model is trained to solve all four tasks simultaneously. Both are evaluated on the newly introduced version of the Fu$\tau$ure dataset, based on a full simulation of the CLD detector concept for the FCC-ee TeraZ run.

    The MultiParTau approach slightly outperforms SingleParTau on the three classification tasks (\tauh identification, decay mode classification, and charge reconstruction) while using approximately one-quarter of the trainable parameters. The relatively small performance differences between the two approaches indicate that both models achieve excellent classification performance, while suggesting that the shared backbone is able to learn representations that are effective across multiple tasks. The only task in which SingleParTau has an advantage is kinematic reconstruction, where it achieves the smallest median $\Delta R$ and the best \ptvis resolution. Nevertheless, the performance gap in kinematic reconstruction remains modest, demonstrating that a unified multi-task architecture can successfully address the full \tauh reconstruction problem.
    
    For \tauh identification, both approaches achieve jet mis-identification rates of $\mathcal{O}(10^{-3})$ at a global average identification efficiency of 80\%. Decay mode classification yields F1 scores of approximately 0.89--0.95 for the dominant decay channels, while charge reconstruction achieves mis-identification rates below $10^{-3}$ up to an efficiency of 80\%, outperforming the QKappa baseline by one to two orders of magnitude. For kinematic reconstruction, SingleParTau achieves median $\Delta R$ values of $3.1$--$3.7\times10^{-3}$ and \ptvis resolutions of 2.51--3.11\%, while MultiParTau achieves comparable performance with median $\Delta R$ values of $3.8$--$5.4\times10^{-3}$ and \ptvis resolutions of 2.54--3.20\%. The original reconstructed jet four-momentum is included as a reconstruction-level cross-check, yielding worse performance than the ML-based models for the considered kinematic observables.

    This work presents the first realistic high-performance ML-based hadronic $\tau$ reconstruction approach for FCC-ee studies, offering \tauh identification, charge reconstruction, decay mode classification, and full visible four-momentum reconstruction. The resulting models provide high-level reconstructed \tauh objects suitable for future FCC-ee analyses, while the MultiParTau approach demonstrates that strong performance across all reconstruction tasks can be achieved with a single unified model.

    Future work will focus on further improving \tauh reconstruction by extending the set of targeted decay modes, including a dedicated class for kaon enriched rare decays, and by incorporating additional information such as secondary vertices. It will also be interesting to study how performance scales with the size of the training dataset through dedicated ablation studies. Finally, extending the scope to include neutrino reconstruction would represent an important step towards a more complete description of \tauh decays.

\section*{Acknowledgments}
    We wish to thank Marko Stamenkovic for fruitful discussions.
    This work has been supported by the Estonian Research Council grants
    PSG864, 
    TARISTU24-TK10, 
    PUTJD1344, 
    PRG2502, 
    Tem-TA23 
    and by the European Regional Development Fund through the CoE program grant TK202. 

\section*{Data availability}
    The dataset used in this paper is made available in Ref.~\cite{dataset}. The software used to produce the results can be found in Ref.~\cite{model_software}.

\section*{Author contributions (CRediT)} 
    \textbf{Nalong-Norman Seeba}: Conceptualization, Methodology, Software, Validation, Formal analysis, Investigation, Data Curation, Writing - Original Draft, Writing - Review \& Editing, Visualization.
    \textbf{Laurits Tani}: Conceptualization, Methodology, Software, Validation, Formal analysis, Investigation, Data Curation, Writing - Original Draft, Writing - Review \& Editing, Visualization, Supervision, Project administration, Funding administration.
    \textbf{Torben Lange}:  Writing - Review \& Editing, Supervision, Funding administration.
    \textbf{Joosep Pata}: Conceptualization, Methodology, Software, Validation, Formal analysis, Investigation, Data Curation, Writing - Original Draft, Writing - Review \& Editing, Visualization, Supervision, Project administration, Funding administration.

\printbibliography

\end{document}